\documentclass[final,sort&compress]{aipproc}
\layoutstyle{6x9}
\usepackage{amssymb,amsmath}
\usepackage{tikz}
\usetikzlibrary{patterns,folding}
\makeatletter
\renewcommand\scriptsize{\@setfontsize\scriptsize{7}{8}}
\makeatother

\begin{document}
\title{A contextual extension of Spekkens' toy model}



\author{Jan-\AA{}ke Larsson}{ address={Institutionen f\"or
    Systemteknik, Link\"opings Universitet, SE-581 83 Link\"oping,
    Sweden} }

\begin{abstract}
  Quantum systems show contextuality. More precisely, it is impossible
  to reproduce the quantum-mechanical predictions using a
  non-contextual realist model, i.e., a model where the outcome of one
  measurement is independent of the choice of compatible measurements
  performed in the measurement context. There has been several
  attempts to quantify the amount of contextuality for specific
  quantum systems, for example, in the number of rays needed in a KS
  proof, or the number of terms in certain inequalities, or in the
  violation, noise sensitivity, and other measures.  This paper is
  about another approach: to use a simple contextual model that
  reproduces the quantum-mechanical contextual behaviour, but not
  necessarily all quantum predictions. The amount of contextuality can
  then be quantified in terms of additional resources needed as
  compared with a similar model without contextuality.  In this case
  the contextual model needs to keep track of the context used, so the
  appropriate measure would be memory. Another way to view this is as
  a memory requirement to be able to reproduce quantum contextuality
  in a realist model.  The model we will use can be viewed as an
  extension of Spekkens' toy model [Phys.\ Rev.\ A \textbf{75}, 032110
  (2007)], and the relation is studied in some detail. To reproduce
  the quantum predictions for the Peres-Mermin square, the memory
  requirement is more than one bit in addition to the memory used for
  the individual outcomes in the corresponding noncontextual model.
\end{abstract}

\maketitle

\section{Introduction}

In this short paper we will see an example of how a noncontextual
model, that attempts to emulate quantum mechanics, is modified into a
contextual model. The larger and more fundamental question
\cite{Einstein1935777}: ``Can the quantum-mechanical description be
considered complete?'' concerns the possibility to augment quantum
mechanics (QM) with additional hidden variables (HVs) that provide a
more detailed, more complete description (see also \cite{VonNeumann31,
  Bohr1935696}).  Attempts to construct such HV models include ones
where, for a given experiment, the observed probability distributions
are used as a HV model \cite{WW01}.  Moreover, there are explicit HV
theories, such as Bohmian mechanics \cite{BH93,Holland93}, which can
reproduce all experiments up to date.  However, these models use
properties that make them less general (using a fundamentally
different model for each experimental setup) or undesirable in other
ways (superluminal influence). Therefore, we need to make additional
assumptions about the structure of the HV model, to avoid this.  The
most famous result in this direction is Bell's theorem \cite{Bell64},
stating that \emph{local} HV models cannot reproduce the QM
correlations between local measurements on some entangled states. That
particular set of additional assumptions will not be studied further
in this paper but we note that this conflict can be tested in
experiment \cite{ADR82,WJSWZ98,RKMSIMW01}.

A second seminal result on HV models reproducing QM predictions is the
Kochen-Specker (KS) theorem \cite{Specker60,Bell66,KS67}. This result
concerns so-called \emph{noncontextual} HV models, in which
measurement outcomes do not depend on which other compatible
measurements are (have been) performed.  There have been several
proposals to test the KS theorem in experiment
\cite{RS93,CG98,SZWZ00,SBZ01,Larsson02,CFRH08,KCBS08,Cabello08,BBCP09},
but there is also a discussion whether the KS theorem can be
experimentally tested at all \cite{Meyer99,Kent99,Mermin99,CK00,
  HKSS01,Appleby02,Cabello02,Breuer02a,Breuer02b,BK04,LaCour09a,CL10}.
We will leave that discussion here, and concentrate on one specific
noncontextual model, Spekkens' toy model \cite{Spekkens07}, which is
known to possess certain QM features but lack others, in particular
contextuality. This paper is concerned with an attempt to add
contextuality to this model, to make it closer to QM, and analyses the
cost of this addition. We will see that the price is to be paid in the
form of a larger ontic state space and a more complicated set of rules
that govern state changes caused by measurement. The paper starts with
a brief description of the system used and the contextuality property
of this QM system, and proceeds with a description of Spekkens' toy
model and the addition needed to make it contextual.

\section{A quantum system and its contextuality}

In this paper, we are going to use a two-qubit system, and
measurements on it. When studying questions about local hidden
variable models, it is common to assume that the system is in a
specific (entangled) state, say the singlet state. Here, we are not
going to assume that the system is in such a state, in fact, what
follows will be valid even if the state is completely mixed.

We will be concerned with measurement outcomes, and what can be
expected to influence these measurement outcomes. Each subsystem can
be measured upon by checking whether the ``spin of the
particle\footnote{Classical intuition translates the systems'
  particle-like behaviour into the belief that we are using particles
  as our systems. This is not true here, but the present discussion
  will side-step this important issue. Already spin-1/2 properties are
  difficult enough, so we will not discuss the more complex issues of
  position and momentum as properties.}'' is up or down in one of the
axis directions.  In quantum-mechanical terms, our four possible
measurement choices (of observables) for each subsystem is
$
\mathbb I,\ \sigma_x,\ \sigma_y,\ \text{or }\sigma_z,
$ 
and these do not commute; for each subsystem we must make one of the
four choices. We denote the outcomes $+1$ for ``up'' and $-1$ for
``down''. We note that there is a difference between doing a separate
measurement of $\sigma_x$ on the two subsystems, and doing a joint
measurement of the product of $\sigma_x$ on both. The first
corresponds to using the operators $
\sigma_x\otimes\mathbb I,\ \text{and}\ \mathbb I\otimes\sigma_x,
$ 
while the second corresponds to the single operator
$
\sigma_x\otimes\sigma_x, $ 
which only outputs one bit of information. We usually assume that this
is the sign of the product of the previous outcomes, but one needs to
be aware that even this innocuous statement is one of
noncontextuality: we assume that the ``spin of the first particle
along the x-axis'' is the same whether we jointly measure the ``spin
of the second particle along the x-axis'' or not.

This is really based in our classical intuition, and the same
(noncontextual) reasoning implies that the outcome for one subsystem
is independent of measurement choice for the second subsystem. Using
classical intuition we expect that the result of measurement of
$\sigma_x\otimes\mathbb I$, should be independent of the choice
between $\mathbb I\otimes\sigma_x$, $\mathbb I\otimes\sigma_y$, or
$\mathbb I\otimes\sigma_z$, as a measurement on the second
subsystem. This classical intuition underlies the notion of
\emph{noncontextuality}: The outcome of the measurement of one
observable should not be influenced by which simultaneous measurement
is performed, which commuting observable that is measured.  The two
outcomes may be dependent, but the outcome of the first measurement
not be influenced by the \emph{choice} of second measurement.

Perhaps it is in place to stress the connection to counterfactual
reasoning: we are saying that ``I have measured
$\sigma_x\otimes\mathbb I$ and $\mathbb I\otimes\sigma_x$, but a
noncontextual model would have given the same outcome for the first in
the pair if I instead would have measured $\sigma_x\otimes\mathbb I$
and $\mathbb I\otimes\sigma_z$''. A classical system could be expected
to obey this, but as we shall see, a quantum system does not.

In this system, one can form the following square of nine observables,
also known as the Peres-Mermin (PM) square \cite{Mermin90, Peres90}
\begin{equation*}
  \left[\begin{matrix}
      \sigma_z \otimes \mathbb I&
      \mathbb I \otimes \sigma_z&
      \sigma_z \otimes \sigma_z\\
      \mathbb I \otimes \sigma_x&
      \sigma_x \otimes \mathbb I&
      \sigma_x \otimes \sigma_x\\
      \sigma_z \otimes \sigma_x&
      \sigma_x \otimes \sigma_z&
      \sigma_y \otimes \sigma_y
\end{matrix}\right].
\end{equation*}
The square is constructed such that the observables within each row
and column commute, corresponding to compatible measurement, and the
product of the operators in any row or column yields $\mathbb I$,
except for the last column, where
\begin{equation*}
  (\sigma_z\otimes\sigma_z)(\sigma_x\otimes\sigma_x)
  (\sigma_y\otimes\sigma_y)=-\mathbb I.
\end{equation*}
In a noncontextual model, the measurement outcomes for each individual
measurement would not depend on the choice of compatible measurement
from the same row or column. A noncontextual model of our system
(containing our two subsystems) would therefore assign $\pm1$ values
to the outcomes within the matrix, and the products of these values
should follow the same rules: the products of the values in any row or
column should be $+1$, except for the last column where the product
should be $-1$. But this is impossible. Under the QM rule, multiplying
the six row and column products would give the value $-1$, but in our
noncontextual model, each individual value appears exactly twice (once
in each row and once in each column) so that the total product must
equal $+1$ (being a product of squares). Therefore, a noncontextual
model must have an even number of $-1$s in the row and column
products, while QM predicts an odd number of $-1$s. A noncontextual
model cannot obey the QM predictions.

\section{Spekkens' toy model}

We will now briefly shift our focus to one explicit noncontextual
model that is specifically designed to mimic QM. It has some of the
properties of QM but lacks others. We will just look at the basic
construction here, for those interested I would recommend reading
\cite{Spekkens07} and references therein.  The toy theory is built on
the following foundational principle (\emph{The knowledge balance
  principle}):

\begin{quote}
  If one has maximal knowledge, then for every system, at every time,
  the amount of knowledge one possesses about the ontic state of the
  system at that time must equal the amount of knowledge one
  lacks \cite{Spekkens07}.
\end{quote}

The model is constructed to mimic QM, so in its simplest form, we want
to construct the equivalent of a qubit. A qubit measurement has two
outcomes, which by the knowledge balance principle means that the
ontic state space of a toy bit has four states. The ontic state space
can be graphically represented as
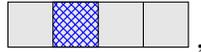
\begin{center}
  \begin{tikzpicture}[scale=.6]
    \draw[fill=black!10] (0,0) -- (4,0) -- (4,1) -- (0,1) -- cycle;%
    \draw[pattern=crosshatch,pattern color=blue] (1,0) -- (2,0) --
    (2,1) -- (1,1) -- cycle;%
    \draw (3,0) -- (3,1);%
  \end{tikzpicture}\ ,
\end{center}
where the ontic state is the second of the four possible. In general,
a state can be a statistical mix of these ontic states.

Our measurement can, at the most, identify two out of these four ontic
states and say: ``the system is in one of these two states'' (as
opposed to being in one of the other two states). Statements like this
identify the epistemic states (the experimentally available outcomes)
of the system. There are six combinations of two ontic states, six
epistemic states. These naturally form three pairs that can be
identified with the three spin-measurement axes\vspace*{-1.5ex}
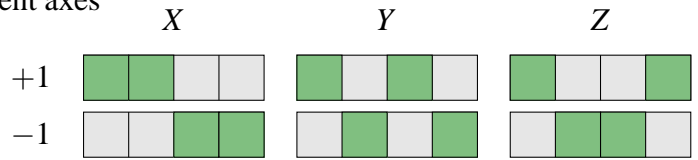
\begin{center}
  \begin{tabular}{c c c c}
    &\raisebox{1ex}{$X$}&\raisebox{1ex}{$Y$}&\raisebox{1ex}{$Z$\phantom{\ .}}\\
    \raisebox{1ex}{$+1$}
    &\begin{tikzpicture}[scale=.6]
      \draw[fill=black!10] (0,0) -- (4,0) -- (4,1) -- (0,1) -- cycle;%
      \draw[fill=black!50!green!50] (0,0) -- (1,0) -- (1,1) -- (0,1) -- cycle;%
      \draw[fill=black!50!green!50] (1,0) -- (2,0) -- (2,1) -- (1,1) -- cycle;%
      \draw (3,0) -- (3,1);%
    \end{tikzpicture}
    &\begin{tikzpicture}[scale=.6]
      \draw[fill=black!10] (0,0) -- (4,0) -- (4,1) -- (0,1) -- cycle;%
      \draw[fill=black!50!green!50] (0,0) -- (1,0) -- (1,1) -- (0,1) -- cycle;%
      \draw[fill=black!50!green!50] (2,0) -- (3,0) -- (3,1) -- (2,1) -- cycle;%
    \end{tikzpicture}
    &\begin{tikzpicture}[scale=.6]
      \draw[fill=black!10] (0,0) -- (4,0) -- (4,1) -- (0,1) -- cycle;%
      \draw[fill=black!50!green!50] (0,0) -- (1,0) -- (1,1) -- (0,1) -- cycle;%
      \draw[fill=black!50!green!50] (3,0) -- (4,0) -- (4,1) -- (3,1) -- cycle;%
      \draw (2,0) -- (2,1);%
    \end{tikzpicture}\phantom{\ .}
    \\
    \raisebox{1ex}{$-1$}
    &\begin{tikzpicture}[scale=.6]
      \draw[fill=black!10] (0,0) -- (4,0) -- (4,1) -- (0,1) -- cycle;%
      \draw[fill=black!50!green!50] (2,0) -- (3,0) -- (3,1) -- (2,1) -- cycle;%
      \draw[fill=black!50!green!50] (3,0) -- (4,0) -- (4,1) -- (3,1) -- cycle;%
      \draw (1,0) -- (1,1);%
    \end{tikzpicture}
    &\begin{tikzpicture}[scale=.6]
      \draw[fill=black!10] (0,0) -- (4,0) -- (4,1) -- (0,1) -- cycle;%
      \draw[fill=black!50!green!50] (3,0) -- (4,0) -- (4,1) -- (3,1) -- cycle;%
      \draw[fill=black!50!green!50] (1,0) -- (2,0) -- (2,1) -- (1,1) -- cycle;%
    \end{tikzpicture}
    &\begin{tikzpicture}[scale=.6]
      \draw[fill=black!10] (0,0) -- (4,0) -- (4,1) -- (0,1) -- cycle;%
      \draw[fill=black!50!green!50] (2,0) -- (3,0) -- (3,1) -- (2,1) -- cycle;%
      \draw[fill=black!50!green!50] (1,0) -- (2,0) -- (2,1) -- (1,1) -- cycle;%
    \end{tikzpicture}\ .
  \end{tabular}\vspace*{-.5ex}
\end{center}
For a system in the ontic state depicted above, a measurement of ``the
spin along the x-axis'' would give the outcome $+1$, because the ontic
state is contained in the corresponding epistemic state. A subsequent
measurement of ``the spin along the y-axis'' would identify the ontic
state completely unless the measurement procedure is allowed to
influence the state. Therefore, the knowledge balance principle forces
the model to include a state change after measurement. The model is
such that the ontic state is randomised after measurement between the
two possibilities in the epistemic state. This removes the possibility
to extract more information about the ontic state than the knowledge
balance principle allows. In this manner, the knowledge balance
principle mimics the uncertainty principle from QM. 

Having presented the single system construction, we will now quickly
review the combination of two such systems into a two-toy-bit
model. The ontic state space now has sixteen points, and a specific
ontic state can be graphically represented as
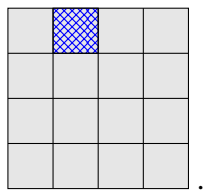
\begin{center}
  \begin{tikzpicture}[scale=.6]
    \draw[fill=black!10] (0,0) -- (4,0) -- (4,4) -- (0,4) -- cycle;%
    \draw[pattern=crosshatch,pattern color=blue] (1,3) -- (2,3) --
    (2,4) -- (1,4) -- cycle;%
    \draw (1,0) -- (1,4);%
    \draw (2,0) -- (2,4);%
    \draw (3,0) -- (3,4);%
    \draw (0,1) -- (4,1);%
    \draw (0,2) -- (4,2);%
    \draw (0,3) -- (4,3);%
  \end{tikzpicture}{\ .}
\end{center}
We can maximally extract two bits of knowledge from this system, and
the knowledge balance principle gives us many epistemic states, for
example
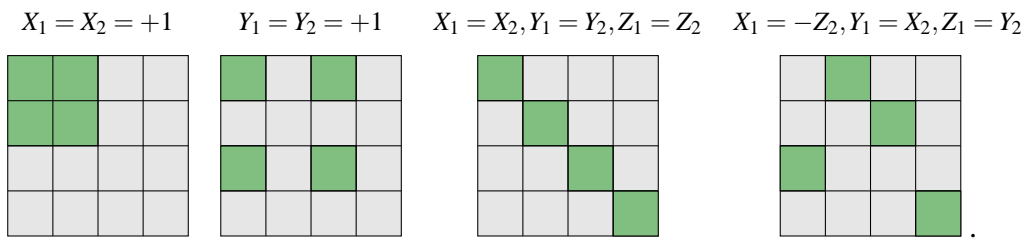
\begin{center}
  \begin{tabular}{c c c c}
    \raisebox{1ex}{\footnotesize$X_1=X_2=+1$}
    &\raisebox{1ex}{\footnotesize$Y_1=Y_2=+1$}
    &\raisebox{1ex}{\footnotesize$X_1=X_2,Y_1=Y_2,Z_1=Z_2$}
    &\raisebox{1ex}{\footnotesize$X_1=-Z_2,Y_1=X_2,Z_1=Y_2$}\\
    \begin{tikzpicture}[scale=.6]
    \draw[fill=black!10] (0,0) -- (4,0) -- (4,4) -- (0,4) -- cycle;%
    \draw[fill=black!50!green!50] (0,2) -- (2,2) -- (2,4) -- (0,4) -- cycle;%
    \draw (1,0) -- (1,4);%
    \draw (2,0) -- (2,4);%
    \draw (3,0) -- (3,4);%
    \draw (0,1) -- (4,1);%
    \draw (0,2) -- (4,2);%
    \draw (0,3) -- (4,3);%
    \end{tikzpicture}
    &\begin{tikzpicture}[scale=.6]
      \draw[fill=black!10] (0,0) -- (4,0) -- (4,4) -- (0,4) -- cycle;%
      \draw[fill=black!50!green!50] (0,1) -- (1,1) -- (1,2) -- (0,2) -- cycle;%
      \draw[fill=black!50!green!50] (0,3) -- (1,3) -- (1,4) -- (0,4) -- cycle;%
      \draw[fill=black!50!green!50] (2,1) -- (3,1) -- (3,2) -- (2,2) -- cycle;%
      \draw[fill=black!50!green!50] (2,3) -- (3,3) -- (3,4) -- (2,4) -- cycle;%
      \draw (1,0) -- (1,4);%
      \draw (2,0) -- (2,4);%
      \draw (3,0) -- (3,4);%
      \draw (0,1) -- (4,1);%
      \draw (0,2) -- (4,2);%
      \draw (0,3) -- (4,3);%
    \end{tikzpicture}
    &\begin{tikzpicture}[scale=.6]
      \draw[fill=black!10] (0,0) -- (4,0) -- (4,4) -- (0,4) -- cycle;%
      \draw[fill=black!50!green!50] (0,3) -- (1,3) -- (1,4) -- (0,4) -- cycle;%
      \draw[fill=black!50!green!50] (1,2) -- (2,2) -- (2,3) -- (1,3) -- cycle;%
      \draw[fill=black!50!green!50] (2,1) -- (3,1) -- (3,2) -- (2,2) -- cycle;%
      \draw[fill=black!50!green!50] (3,0) -- (4,0) -- (4,1) -- (3,1) -- cycle;%
      \draw (1,0) -- (1,4);%
      \draw (2,0) -- (2,4);%
      \draw (3,0) -- (3,4);%
      \draw (0,1) -- (4,1);%
      \draw (0,2) -- (4,2);%
      \draw (0,3) -- (4,3);%
    \end{tikzpicture}
    &\begin{tikzpicture}[scale=.6]
      \draw[fill=black!10] (0,0) -- (4,0) -- (4,4) -- (0,4) -- cycle;%
      \draw[fill=black!50!green!50] (0,1) -- (1,1) -- (1,2) -- (0,2) -- cycle;%
      \draw[fill=black!50!green!50] (1,3) -- (2,3) -- (2,4) -- (1,4) -- cycle;%
      \draw[fill=black!50!green!50] (2,2) -- (3,2) -- (3,3) -- (2,3) -- cycle;%
      \draw[fill=black!50!green!50] (3,0) -- (4,0) -- (4,1) -- (3,1) -- cycle;%
      \draw (1,0) -- (1,4);%
      \draw (2,0) -- (2,4);%
      \draw (3,0) -- (3,4);%
      \draw (0,1) -- (4,1);%
      \draw (0,2) -- (4,2);%
      \draw (0,3) -- (4,3);%
    \end{tikzpicture}\ .
  \end{tabular}
\end{center}
A measurement determining which epistemic state the system is in again
randomises between the four ontic states contained in the epistemic
state.  Note that not all combinations of four ontic states are
allowed as epistemic states, again for details, see
\cite{Spekkens07}. This is sufficiently close to two-qubit systems in
QM to give several behaviours that can be used to mimic quantum-like
behaviour. A partial list would include Noncommutativity
(uncertainty), Interference, Remote steering, No cloning, No
broadcasting, Mutually Unbiased Partitions, Dense coding, Entanglement
monogamy, Teleportation, Positive Operator Valued Measures, \ldots .

A subsystem measurement on a two-toy-bit model would extract less than
maximal information, and the corresponding epistemic states contain
eight ontic states. The ontic state of the measured subsystem is
randomised, while the ontic state of the unmeasured subsystem is not.
In fact, whenever nonmaximal information is extracted, the ontic state
is not maximally randomised within the epistemic state, but it is
easiest to understand in the example of a subsystem measurement, as in
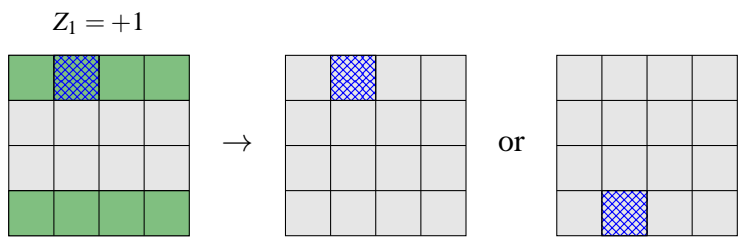
\begin{center}
  \begin{tabular}{c c c c c}
    \raisebox{1ex}{\footnotesize$Z_1=+1$}\\
    \begin{tikzpicture}[scale=.6]
      \draw[fill=black!10] (0,0) -- (4,0) -- (4,4) -- (0,4) -- cycle;%
      \draw[fill=black!50!green!50] (0,0) -- (4,0) -- (4,1) -- (0,1)
      -- cycle;%
      \draw[fill=black!50!green!50] (0,3) -- (4,3) -- (4,4) -- (0,4)
      -- cycle;%
      \draw[pattern=crosshatch,pattern color=blue] (1,3) -- (2,3) --
      (2,4) -- (1,4) -- cycle;%
      \draw (1,0) -- (1,4);%
      \draw (2,0) -- (2,4);%
      \draw (3,0) -- (3,4);%
      \draw (0,1) -- (4,1);%
      \draw (0,2) -- (4,2);%
      \draw (0,3) -- (4,3);%
    \end{tikzpicture}
    &\raisebox{2.7em}{$\rightarrow$}&
    \begin{tikzpicture}[scale=.6]
      \draw[fill=black!10] (0,0) -- (4,0) -- (4,4) -- (0,4) -- cycle;%
      \draw[pattern=crosshatch,pattern color=blue] (1,3) -- (2,3) --
      (2,4) -- (1,4) -- cycle;%
      \draw (1,0) -- (1,4);%
      \draw (2,0) -- (2,4);%
      \draw (3,0) -- (3,4);%
      \draw (0,1) -- (4,1);%
      \draw (0,2) -- (4,2);%
      \draw (0,3) -- (4,3);%
    \end{tikzpicture}
    &\raisebox{2.7em}{or}&
    \begin{tikzpicture}[scale=.6]
    \draw[fill=black!10] (0,0) -- (4,0) -- (4,4) -- (0,4) -- cycle;%
    \draw[pattern=crosshatch,pattern color=blue] (1,0) -- (2,0) -- (2,1) -- (1,1) -- cycle;%
    \draw (1,0) -- (1,4);%
    \draw (2,0) -- (2,4);%
    \draw (3,0) -- (3,4);%
    \draw (0,1) -- (4,1);%
    \draw (0,2) -- (4,2);%
    \draw (0,3) -- (4,3);%
    \end{tikzpicture}\ .
  \end{tabular}
\end{center}
An example of a protocol that uses a two-qu/toy-bit system and
performs transformations on one subsystem by itself is dense coding;
we can observe that the ontic state space of a one-toy-bit subsystem
contains exactly four states, which is just enough to allow for dense
coding.

However, not all QM behaviour is captured. The most important missing
features are Contextuality, Nonlocality, Continuum of states, Two
levels of a three-level system is not a two-level system, ``Quantum''
computational speedup, \ldots; we will concentrate on the first
item. It is simple to see that the model is noncontextual, because the
measurement procedure is such that an outcome of a joint measurement
outcome is given by the product of the individual outcomes. Our
example ontic state would give the following results:
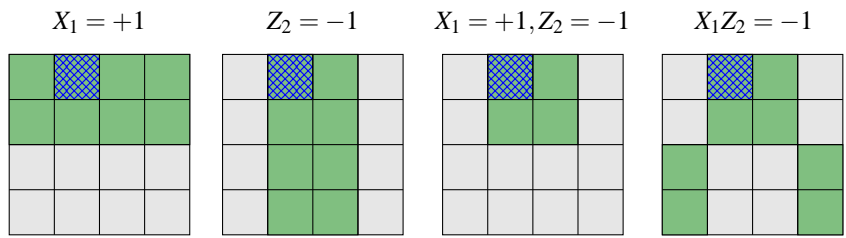
\begin{center}
  \begin{tabular}{c c c c}
    \raisebox{1ex}{\footnotesize$X_1=+1$}
    &\raisebox{1ex}{\footnotesize$Z_2=-1$}
    &\raisebox{1ex}{\footnotesize$X_1=+1,Z_2=-1$}
    &\raisebox{1ex}{\footnotesize$X_1Z_2=-1$\phantom{\ .}}\\
    \begin{tikzpicture}[scale=.6]
      \draw[fill=black!10] (0,0) -- (4,0) -- (4,4) -- (0,4) -- cycle;%
      \draw[fill=black!50!green!50] (0,2) -- (4,2) -- (4,4) -- (0,4)
      -- cycle;%
      \draw[pattern=crosshatch,pattern color=blue] (1,3) -- (2,3) --
      (2,4) -- (1,4) -- cycle;%
      \draw (1,0) -- (1,4);%
      \draw (2,0) -- (2,4);%
      \draw (3,0) -- (3,4);%
      \draw (0,1) -- (4,1);%
      \draw (0,2) -- (4,2);%
      \draw (0,3) -- (4,3);%
    \end{tikzpicture}
    &\begin{tikzpicture}[scale=.6]
      \draw[fill=black!10] (0,0) -- (4,0) -- (4,4) -- (0,4) -- cycle;%
      \draw[fill=black!50!green!50] (1,0) -- (3,0) -- (3,4) -- (1,4) -- cycle;%
      \draw[pattern=crosshatch,pattern color=blue] (1,3) -- (2,3) -- (2,4) -- (1,4) -- cycle;%
      \draw (1,0) -- (1,4);%
      \draw (2,0) -- (2,4);%
      \draw (3,0) -- (3,4);%
      \draw (0,1) -- (4,1);%
      \draw (0,2) -- (4,2);%
      \draw (0,3) -- (4,3);%
    \end{tikzpicture}
    &\begin{tikzpicture}[scale=.6]
      \draw[fill=black!10] (0,0) -- (4,0) -- (4,4) -- (0,4) -- cycle;%
      \draw[fill=black!50!green!50] (1,2) -- (3,2) -- (3,4) -- (1,4) -- cycle;%
      \draw[pattern=crosshatch,pattern color=blue] (1,3) -- (2,3) -- (2,4) -- (1,4) -- cycle;%
      \draw (1,0) -- (1,4);%
      \draw (2,0) -- (2,4);%
      \draw (3,0) -- (3,4);%
      \draw (0,1) -- (4,1);%
      \draw (0,2) -- (4,2);%
      \draw (0,3) -- (4,3);%
  \end{tikzpicture}
  &\begin{tikzpicture}[scale=.6]
    \draw[fill=black!10] (0,0) -- (4,0) -- (4,4) -- (0,4) -- cycle;%
    \draw[fill=black!50!green!50] (1,2) -- (3,2) -- (3,4) -- (1,4) -- cycle;%
    \draw[fill=black!50!green!50] (0,0) -- (1,0) -- (1,2) -- (0,2) -- cycle;%
    \draw[fill=black!50!green!50] (3,0) -- (4,0) -- (4,2) -- (3,2) -- cycle;%
    \draw[pattern=crosshatch,pattern color=blue] (1,3) -- (2,3) -- (2,4) -- (1,4) -- cycle;%
    \draw (1,0) -- (1,4);%
    \draw (2,0) -- (2,4);%
    \draw (3,0) -- (3,4);%
    \draw (0,1) -- (4,1);%
    \draw (0,2) -- (4,2);%
    \draw (0,3) -- (4,3);%
    \end{tikzpicture}\ .
  \end{tabular}
\end{center}
The measurement outcome $X_1Z_2=-1$ is one of the outcomes of the PM
square, when used on the toy model. For this particular ontic state,
writing out all the combinations results in
\begin{equation*}
  \left[\begin{matrix}
      Z_1&Z_2&Z_1Z_2\\
      X_2&X_1&X_1X_2\\
      Z_1X_2&X_1Z_2&Y_1Y_2
    \end{matrix}\right]
  =
  \left[\begin{matrix}
      +1&-1&-1\\
      +1&+1&+1\\
      +1&-1&-1
    \end{matrix}\right].
\end{equation*}
Writing out the signs contained in the PM square for each of the ontic
states results in
\begin{center}
  \begin{tikzpicture}[scale=1 ,align=center] 
  \scriptsize
    \draw[fill=black!10] (0,0) -- (4,0) -- (4,4) -- (0,4) -- cycle;%
    \draw (1,0) -- (1,4);%
    \draw (2,0) -- (2,4);%
    \draw (3,0) -- (3,4);%
    \draw (0,1) -- (4,1);%
    \draw (0,2) -- (4,2);%
    \draw (0,3) -- (4,3);%
    \draw (0.5,3.5) node {$+++$\\$+++$\\$+++$};%
    \draw (0.5,2.5) node {$-+-$\\$+++$\\$-+-$};%
    \draw (0.5,1.5) node {$-+-$\\$+--$\\$--+$};%
    \draw (0.5,0.5) node {$+++$\\$+--$\\$+--$};%
    \draw (1.5,3.5) node {$+--$\\$+++$\\$+--$};%
    \draw (1.5,2.5) node {$--+$\\$+++$\\$--+$};%
    \draw (1.5,1.5) node {$--+$\\$+--$\\$-+-$};%
    \draw (1.5,0.5) node {$+--$\\$+--$\\$+++$};%
    \draw (2.5,3.5) node {$+--$\\$-+-$\\$--+$};%
    \draw (2.5,2.5) node {$--+$\\$-+-$\\$+--$};%
    \draw (2.5,1.5) node {$--+$\\$--+$\\$+++$};%
    \draw (2.5,0.5) node {$+--$\\$--+$\\$-+-$};%
    \draw (3.5,3.5) node {$+++$\\$-+-$\\$-+-$};%
    \draw (3.5,2.5) node {$-+-$\\$-+-$\\$+++$};%
    \draw (3.5,1.5) node {$-+-$\\$--+$\\$+--$};%
    \draw (3.5,0.5) node {$+++$\\$--+$\\$--+$};%
  \end{tikzpicture}\ .
\end{center}
Here, all rows and columns have the product $+1$, so that the ontic
states can be indexed by the top left four signs of the PM-square
values. Also, the last column does not fulfil the QM prediction, a
product equal to $-1$.
 
This can be viewed as a ``finite state machine'' (specifically, a
``stochastic Mealy machine'') as follows: each ontic state is one of
the sixteen possible states of the machine. When a measurement is
performed, the machine outputs a $\pm1$ value that depends on its
ontic state and the chosen measurement (the ``input'' to the machine),
and then changes state. In the toy model this state change is to a
random ontic state within the epistemic state.  Below you can see an
example, where the large discs are two (ontic) states of the finite
state machine and the arrows are state transitions. We start in the
left ontic state, perform a measurement of $Z_1$ which outputs $+1$
and changes state randomly to either the same state or a new state
with opposite values assigned to the non-commuting observables. The
contradiction to the QM prediction is indicated with a dotted ellipse,
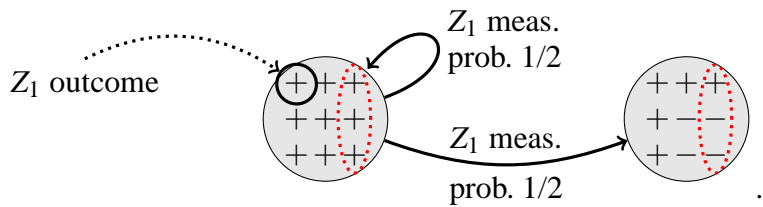
\begin{center}
  \begin{tikzpicture}[scale=.8,align=center] 
    \draw[use as bounding box] (-1,-1.3) (7,1.8);%
    \draw (0,0) node(a)[draw,circle, fill=black!10,inner sep=0pt]
    {$+++$\\$+++$\\$+++$};%
    \draw[red,dotted,very thick] (a) +(.48,0) circle[x radius=.27,y
    radius=.9];%
    \draw (6,0) node(b)[draw,circle,fill=black!10,inner sep=0pt]
    {$+++$\\$+--$\\$+--$};%
    \draw[red,dotted,very thick] (b) +(.48,0) circle[x radius=.27,y
    radius=.9];%
    \draw (-.48,.58) node[draw,very thick,circle,minimum size=15](c)
    {};%
    \draw[->,very thick] (a) to[bend right=20] node[above]{$Z_1$ meas.}
    node[below]{prob. 1/2} (b);%
    \draw[->,very thick] (a) edge[in=50,out=20,loop]
    node[right,align=left]{ $Z_1$ meas.\\prob.\ 1/2 } (a);%
    \draw[->,dotted,very thick] (-4,1) node[below] {$Z_1$ outcome} to[bend
    left] (c);%
  \end{tikzpicture}\ .
\end{center}
This description is mathematically equivalent to the previous. Of
course, the above diagram is only a small part of the state space and
transitions, since there are sixteen different ontic states and nine
possible measurement choices, all corresponding to a different network
of state transitions.  Using the language of finite state machines, we
can now add properties to the model that were previously difficult to
conceptualise. One less desirable effect is that the epistemic states
are more difficult to present within the description. But, since our
aim is to add properties, we will use the state machine formalism
below.

\section{Adding contextuality to the toy model}

We now want to extend the model so that it gives the QM predictions,
including the PM square results. For brevity, we will restrict
ourselves to sequential measurements \cite{Larsson2011401}, and a more
complete presentation of the procedure can be found in \cite{KGPLC}.
To extend the model properly, we will need to add (ontic) states to
the finite state machine. The basic questions are what states we add,
how many, and if there is a minimal extension. The leftmost state
below is already present, and the two others are example states that
are possible to add:
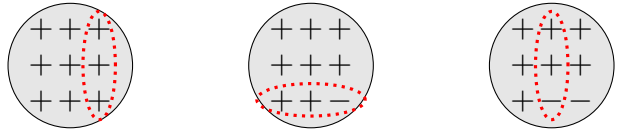
\begin{center}
  \begin{tikzpicture}[scale=.8,align=center] 
    \draw (0,0) node(a)[draw,circle, fill=black!10,inner sep=0pt]
    {\color{black} $+++$\\$+++$\\$+++$};%
    \draw[red,dotted,very thick] (a) +(.48,0) circle[x radius=.27,y
    radius=.9];%
    \draw (4,0) node(b)[draw,circle, fill=black!10,inner sep=0pt]
    {\color{black} $+++$\\$+++$\\$++-$};%
    \draw[red,dotted,very thick] (b) +(0,-.57) circle[x radius=.9,y
    radius=.27];%
    \draw (8,0) node(a)[draw,circle, fill=black!10,inner sep=0pt]
    {\color{black} $+++$\\$+++$\\$+--$};%
    \draw[red,dotted,very thick] (a) +(0,0) circle[x radius=.27,y
    radius=.9];%
  \end{tikzpicture}\ .
\end{center}
Our starting point will be Spekkens' model, where state transitions
always keep the values assigned to commuting observables constant, and
randomise the values assigned to non-commuting observables. Our new
larger model cannot obey this strictly (although it will preserve the
value assigned to the measured observable). This is because of our
desire to reproduce the QM predictions for the last column. To do
this, some transitions induced by measurement in the last column
\emph{must} change one of the values in the column. Indeed, this
transition must go to a state for which the rightmost column product
is $-1$, as in the below example (we will suppress the randomisation
of values of non-commuting observables in what follows)
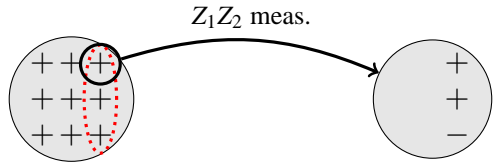
\begin{center}
  \begin{tikzpicture}[scale=.8,align=center] 
    \draw[use as bounding box] (-1,-1) (7,1.4);%
    \draw (0,0) node(a)[draw,circle, fill=black!10,inner sep=0pt]
    {$+++$\\$+++$\\$+++$};%
    \draw[red,dotted,very thick] (a) +(.48,0) circle[x radius=.27,y
    radius=.9];%
    \draw (6,0) node(b)[draw,circle,fill=black!10,inner sep=0pt]
    {$\phantom{++}+$\\$\phantom{++}+$\\$\phantom{++}-$};%
    \draw (.48,.58) node[draw,very thick,circle,minimum size=15](c)
    {};%
    \draw[->,very thick] (c) to[bend left=20]
    node[above]{\footnotesize $Z_1Z_2$ meas.}  (b);%
  \end{tikzpicture}\ .
\end{center}

Thus, we will need to add ontic states of the type indicated above.
However, it turns out that \emph{one} transition of this type from the
rightmost column is not enough. If, in the above example, we were to
perform measurements of the right column observables in the order from
the bottom one to the top one, the state change is too late to stop us
from finding the column product $+1$. Therefore, at least \emph{two}
measurements in the column must give rise to state transitions, both
leading to a state with no contradiction in the last column (but a
contradiction elsewhere)
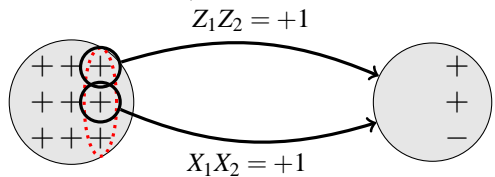
\begin{center}
  \begin{tikzpicture}[scale=.8,align=center] 
    \draw[use as bounding box] (-1,-1) (7,1.4);%
    \draw (0,0) node(a)[draw,circle, fill=black!10,inner sep=0pt]
    {$+++$\\$+++$\\$+++$};%
    \draw[red,dotted,very thick] (a) +(.48,0) circle[x radius=.27,y
    radius=.9];%
    \draw (6,0) node(b)[draw,circle,fill=black!10,inner sep=0pt]
    {$\phantom{++}+$\\$\phantom{++}+$\\$\phantom{++}-$};%
    \draw (.48,.58) node[draw,very thick,circle,minimum size=15](c)
    {};%
    \draw (.48,0) node[draw,very thick,circle,minimum size=15](d)
    {};%
    \draw[->,very thick] (c) to[bend left=20]
    node[above]{\footnotesize$Z_1Z_2=+1$} (b);%
    \draw[->,very thick] (d) to[bend right=20]
    node[below]{\footnotesize$X_1X_2=+1$} (b);%
  \end{tikzpicture}\ .
\end{center}

Furthermore, the two transitions must go to \emph{different} states,
because otherwise we will find the column product $+1$ since the state
transition must preserve the value assigned to the measured observable
(again, this can be demonstrated by measuring in the order from bottom
to top above). 
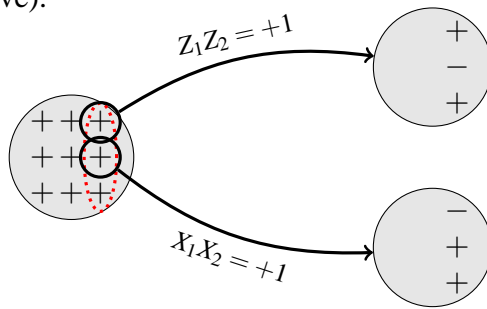
\begin{center}
  \begin{tikzpicture}[scale=.8,align=center] 
    \draw[use as bounding box] (-1,-2.5) (7,2.1);%
    \draw (0,0) node(a)[draw,circle, fill=black!10,inner sep=0pt]
    {$+++$\\$+++$\\$+++$};%
    \draw[red,dotted,very thick] (a) +(.48,0) circle[x radius=.27,y
    radius=.9];%
    \draw (6,1.5) node(b)[draw,circle,fill=black!10,inner sep=0pt]
    {$\phantom{++}+$\\$\phantom{++}-$\\$\phantom{++}+$};%
    \draw (6,-1.5) node(c)[draw,circle,fill=black!10,inner sep=0pt]
    {$\phantom{++}-$\\$\phantom{++}+$\\$\phantom{++}+$};%
    \draw[->,very thick] (.48,.58) node(b1)[draw,very
    thick,circle,minimum size=15] {} (b1) to[bend left=20]
    node[sloped,above]{\footnotesize$Z_1Z_2=+1$} (b);%
    \draw[->,very thick] (.48,0) node(c1)[draw,very
    thick,circle,minimum size=15] {} (c1) to[bend right=20]
    node[sloped,below]{\footnotesize$X_1X_2=+1$} (c);%
  \end{tikzpicture}\ .
\end{center}

There are now a few alternatives on how to continue to fill in values
in the new states. Note that each ontic state still needs to have an
even number of $-1$ row and column products (an assignment of outcomes
in a single ontic state must be noncontextual). Therefore, we also
need to decide where the $-1$ product should go among the five
alternatives. Some alternatives can be ruled out by using appropriate
sequences of measurement choices, but the somewhat tedious procedure
\cite{KGPLC} will not be repeated here. As it turns out, there is a
model with four states that reproduces the QM predictions for the PM
square (in the sense that the products are the expected ones)
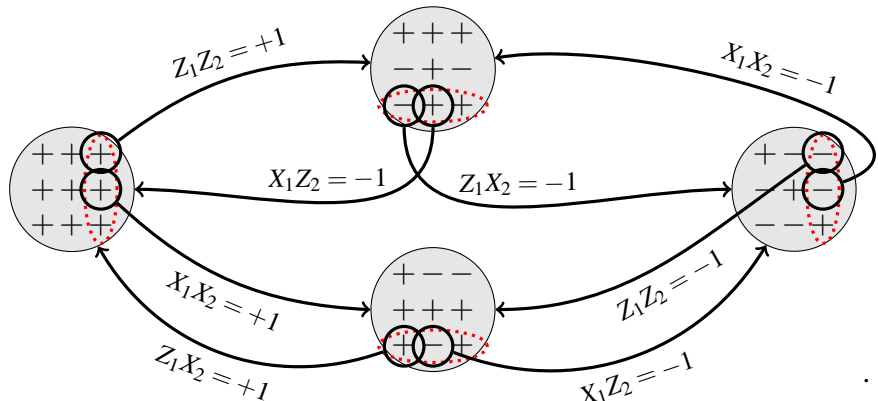
\begin{center}
  \begin{tikzpicture}[scale=.8,align=center] 
    \draw[use as bounding box] (-1,-3.2) (13,3);%
    \draw (0,0) node(a)[draw,circle, fill=black!10,inner sep=0pt]
    {$+++$\\$+++$\\$+++$};%
    \draw[red,dotted,very thick] (a) +(.48,0) circle[x radius=.27,y
    radius=.9];%
    \draw (6,2) node(b)[draw,circle,fill=black!10,inner sep=0pt]
    {$+++$\\$-+-$\\$-++$};%
    \draw[red,dotted,very thick] (b) +(0,-.61) circle[x radius=.9,y
    radius=.27];%
    \draw (6,-2) node(c)[draw,circle,fill=black!10,inner sep=0pt]
    {$+--$\\$+++$\\$+-+$};%
    \draw[red,dotted,very thick] (c) +(0,-.61) circle[x radius=.9,y
    radius=.27];%
    \draw (12,0) node(d)[draw,circle, fill=black!10,inner sep=0pt]
    {$+--$\\$-+-$\\$--+$};%
    \draw[red,dotted,very thick] (d) +(.48,0) circle[x radius=.27,y
    radius=.9];%
    \draw[->,very thick](a)+(.48,.61) node(a1)[draw,very
    thick,circle,minimum size=15] {} (a1) to[bend left=20]
    node[sloped,above]{\footnotesize$Z_1Z_2=+1$} (b);%
    \draw[->,very thick](a)+(.48,0) node(a2)[draw,very
    thick,circle,minimum size=15] {} (a2) to[bend right=20]
    node[sloped,below]{\footnotesize$X_1X_2=+1$} (c);%
    \draw[->,very thick](b)+(-.48,-.61) node(b1)[draw,very
    thick,circle,minimum size=15] {} (b1) to[out=270,in=180]
    node[sloped,above]{\footnotesize$Z_1X_2=-1$} (d);%
    \draw[->,very thick](b)+(0,-.61) node(b2)[draw,very
    thick,circle,minimum size=15] {} (b2) to[out=270,in=0]
    node[sloped,above]{\footnotesize$X_1Z_2=-1$} (a);%
    \draw[->,very thick](c)+(-.48,-.61) node(c1)[draw,very
    thick,circle,minimum size=15] {} (c1) to[bend left=40]
    node[sloped,below]{\footnotesize$Z_1X_2=+1$} (a);%
    \draw[->,very thick](c)+(0,-.61) node(c2)[draw,very
    thick,circle,minimum size=15] {} (c2) to[bend right=40]
    node[sloped,below]{\footnotesize$X_1Z_2=-1$} (d);%
    \draw[->,very thick](d)+(.48,.61) node(d1)[draw,very
    thick,circle,minimum size=15] {} (d1) to[out=215,in=0]
    node[sloped,below]{\footnotesize$Z_1Z_2=-1$} (c);%
    \draw[->,very thick](d)+(.48,0) node(d2)[draw,very
    thick,circle,minimum size=15] {} (d2) to[out=20,in=10]
    node[sloped,above]{\footnotesize$X_1X_2=-1$} (b);%
  \end{tikzpicture}\ .
\end{center}

We have added three ontic states to the one we started with, so it is
tempting to conclude that the model would need 64 ontic states instead
of 16. But a close inspection reveals that the rightmost state above
already is present in Spekkens' original model. Further, the two
middle states both have their contradiction in the last row, and form
a similar class of ontic states. Thus, we only need to add 16 states
to Spekkens' model. Our new ontic state space can be depicted by
repeating the original table beside another where the bottom right
sign has been inverted. The above four-state machine has also been
delineated below.
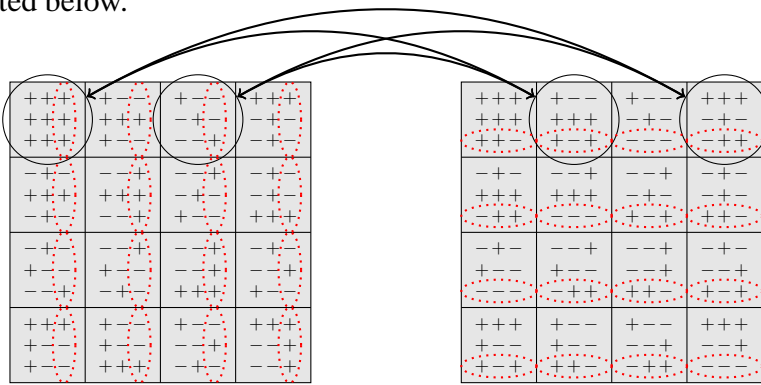
\begin{center}
  \begin{tikzpicture}[scale=1,align=center] 
    \scriptsize
    \draw[use as bounding box] (0,0) (10,4.7);%
    \draw[fill=black!10] (0,0) -- (4,0) -- (4,4) -- (0,4) -- cycle;%
    \draw (1,0) -- (1,4);%
    \draw (2,0) -- (2,4);%
    \draw (3,0) -- (3,4);%
    \draw (0,1) -- (4,1);%
    \draw (0,2) -- (4,2);%
    \draw (0,3) -- (4,3);%
    \draw (0.5,3.5) node(a)[draw,circle] {$+++$\\$+++$\\$+++$};%
    \draw (0.5,2.5) node {$-+-$\\$+++$\\$-+-$};%
    \draw (0.5,1.5) node {$-+-$\\$+--$\\$--+$};%
    \draw (0.5,0.5) node {$+++$\\$+--$\\$+--$};%
    \draw (1.5,3.5) node {$+--$\\$+++$\\$+--$};%
    \draw (1.5,2.5) node {$--+$\\$+++$\\$--+$};%
    \draw (1.5,1.5) node {$--+$\\$+--$\\$-+-$};%
    \draw (1.5,0.5) node {$+--$\\$+--$\\$+++$};%
    \draw (2.5,3.5) node(d)[draw,circle] {$+--$\\$-+-$\\$--+$};%
    \draw (2.5,2.5) node {$--+$\\$-+-$\\$+--$};%
    \draw (2.5,1.5) node {$--+$\\$--+$\\$+++$};%
    \draw (2.5,0.5) node {$+--$\\$--+$\\$-+-$};%
    \draw (3.5,3.5) node {$+++$\\$-+-$\\$-+-$};%
    \draw (3.5,2.5) node {$-+-$\\$-+-$\\$+++$};%
    \draw (3.5,1.5) node {$-+-$\\$--+$\\$+--$};%
    \draw (3.5,0.5) node {$+++$\\$--+$\\$--+$};%

    \draw[fill=black!10] (6,0) -- (10,0) -- (10,4) -- (6,4) -- cycle;%
    \draw (7,0) -- (7,4);%
    \draw (8,0) -- (8,4);%
    \draw (9,0) -- (9,4);%
    \draw (6,1) -- (10,1);%
    \draw (6,2) -- (10,2);%
    \draw (6,3) -- (10,3);%
    \draw (6.5,3.5) node {$+++$\\$+++$\\$++-$};%
    \draw (6.5,2.5) node {$-+-$\\$+++$\\$-++$};%
    \draw (6.5,1.5) node {$-+-$\\$+--$\\$---$};%
    \draw (6.5,0.5) node {$+++$\\$+--$\\$+-+$};%
    \draw (7.5,3.5) node(c)[draw,circle] {$+--$\\$+++$\\$+-+$};%
    \draw (7.5,2.5) node {$--+$\\$+++$\\$---$};%
    \draw (7.5,1.5) node {$--+$\\$+--$\\$-++$};%
    \draw (7.5,0.5) node {$+--$\\$+--$\\$++-$};%
    \draw (8.5,3.5) node {$+--$\\$-+-$\\$---$};%
    \draw (8.5,2.5) node {$--+$\\$-+-$\\$+-+$};%
    \draw (8.5,1.5) node {$--+$\\$--+$\\$++-$};%
    \draw (8.5,0.5) node {$+--$\\$--+$\\$-++$};%
    \draw (9.5,3.5) node(b)[draw,circle] {$+++$\\$-+-$\\$-++$};%
    \draw (9.5,2.5) node {$-+-$\\$-+-$\\$++-$};%
    \draw (9.5,1.5) node {$-+-$\\$--+$\\$+-+$};%
    \draw (9.5,0.5) node {$+++$\\$--+$\\$---$};%

    \draw[thick,<->] (a) to[bend left=30] (c);%
    \draw[thick,<->] (a) to[bend left=30] (b);%
    \draw[thick,<->] (d) to[bend left=30] (c);%
    \draw[thick,<->] (d) to[bend left=30] (b);%

    \draw[red,dotted,thick] (0.5,3.5) +(.22,0) circle[x radius=.15,y
    radius=.5];%
    \draw[red,dotted,thick] (0.5,2.5) +(.22,0) circle[x radius=.15,y
    radius=.5];%
    \draw[red,dotted,thick] (0.5,1.5) +(.22,0) circle[x radius=.15,y
    radius=.5];%
    \draw[red,dotted,thick] (0.5,0.5) +(.22,0) circle[x radius=.15,y
    radius=.5];%
    \draw[red,dotted,thick] (1.5,3.5) +(.22,0) circle[x radius=.15,y
    radius=.5];%
    \draw[red,dotted,thick] (1.5,2.5) +(.22,0) circle[x radius=.15,y
    radius=.5];%
    \draw[red,dotted,thick] (1.5,1.5) +(.22,0) circle[x radius=.15,y
    radius=.5];%
    \draw[red,dotted,thick] (1.5,0.5) +(.22,0) circle[x radius=.15,y
    radius=.5];%
    \draw[red,dotted,thick] (2.5,3.5) +(.22,0) circle[x radius=.15,y
    radius=.5];%
    \draw[red,dotted,thick] (2.5,2.5) +(.22,0) circle[x radius=.15,y
    radius=.5];%
    \draw[red,dotted,thick] (2.5,1.5) +(.22,0) circle[x radius=.15,y
    radius=.5];%
    \draw[red,dotted,thick] (2.5,0.5) +(.22,0) circle[x radius=.15,y
    radius=.5];%
    \draw[red,dotted,thick] (3.5,3.5) +(.22,0) circle[x radius=.15,y
    radius=.5];%
    \draw[red,dotted,thick] (3.5,2.5) +(.22,0) circle[x radius=.15,y
    radius=.5];%
    \draw[red,dotted,thick] (3.5,1.5) +(.22,0) circle[x radius=.15,y
    radius=.5];%
    \draw[red,dotted,thick] (3.5,0.5) +(.22,0) circle[x radius=.15,y
    radius=.5];%

    \draw[red,dotted,thick] (6.5,3.5) +(0,-.28) circle[x radius=.5,y
    radius=.15];%
    \draw[red,dotted,thick] (6.5,2.5) +(0,-.28) circle[x radius=.5,y
    radius=.15];%
    \draw[red,dotted,thick] (6.5,1.5) +(0,-.28) circle[x radius=.5,y
    radius=.15];%
    \draw[red,dotted,thick] (6.5,0.5) +(0,-.28) circle[x radius=.5,y
    radius=.15];%
    \draw[red,dotted,thick] (7.5,3.5) +(0,-.28) circle[x radius=.5,y
    radius=.15];%
    \draw[red,dotted,thick] (7.5,2.5) +(0,-.28) circle[x radius=.5,y
    radius=.15];%
    \draw[red,dotted,thick] (7.5,1.5) +(0,-.28) circle[x radius=.5,y
    radius=.15];%
    \draw[red,dotted,thick] (7.5,0.5) +(0,-.28) circle[x radius=.5,y
    radius=.15];%
    \draw[red,dotted,thick] (8.5,3.5) +(0,-.28) circle[x radius=.5,y
    radius=.15];%
    \draw[red,dotted,thick] (8.5,2.5) +(0,-.28) circle[x radius=.5,y
    radius=.15];%
    \draw[red,dotted,thick] (8.5,1.5) +(0,-.28) circle[x radius=.5,y
    radius=.15];%
    \draw[red,dotted,thick] (8.5,0.5) +(0,-.28) circle[x radius=.5,y
    radius=.15];%
    \draw[red,dotted,thick] (9.5,3.5) +(0,-.28) circle[x radius=.5,y
    radius=.15];%
    \draw[red,dotted,thick] (9.5,2.5) +(0,-.28) circle[x radius=.5,y
    radius=.15];%
    \draw[red,dotted,thick] (9.5,1.5) +(0,-.28) circle[x radius=.5,y
    radius=.15];%
    \draw[red,dotted,thick] (9.5,0.5) +(0,-.28) circle[x radius=.5,y
    radius=.15];%
  \end{tikzpicture}\ .
\end{center}
It is a simple matter to include the randomisation of the
non-compatible measurement outcomes, but we will skip the details in
this short note.

\section{Conclusions}

Quantum Contextuality is one of the more puzzling properties of QM, so
any tool that helps us understand contextuality, ultimately helps us
to understand QM. Spekkens' toy model is helpful in many ways because
it provides a middle ground where we can describe and discuss many of
the phenomena associated with QM and Quantum Information Theory,
without going to the full-blown QM formalism.

This paper attempts to extend this middle ground to include quantum
contextuality into the description and discussion.  Using the language
of finite state machines, we have successfully included
quantum-contextual behaviour. The present note is limited to the case
of sequential measurements on pairs of ``toy'' bits (resembling
qubits), and restricts the measurement choices to (the equivalent of)
``spin measurements'' along two axes.  Unfortunately, the lacking
third axis means that the model is incomplete, and it seems that the
state space needs to be extended even further to incorporate this
\cite{KGPLC}. Adding a third measurement also adds many relations
between the available measurement choices, and results in ten
different PM squares \cite{Cabello10}, instead of one.

Nevertheless, we have seen how to extend Spekkens' toy model to
include two more items in the list of quantum-like behaviours that it
mimics: Contextuality, and Nonlocality. The latter is included because
a two-toy-bit system with separated subsystems would need nonlocal
influence to show the needed behaviour: \emph{both} measurement
choices are needed to determine the state transition in the model. One
property that is certainly still missing is a continuum of states, but
we have yet to see whether two levels of a three-level system is a two
level system or not. We we have not yet seen the three-level version
of this new model, so we have nothing to compare with. The possibility
of a ``Quantum'' computational speedup is also still open.

There is a price to pay for this extension, however: the knowledge
balance principle fails. There are more than twice as many ontic
states as there are epistemic states. Measured in (ordinary) bits of
information, the model uses one additional bit of information that
keeps track of where the contradiction is in the PM square (last row
or last column), and this bit is not accessible to the
experimentalist. One way of interpreting this is that the system keeps
an internal memory of what the experimenter previously has done to the
system.  This one bit of information can be proved to be a lower bound
for models that obey the QM predictions \cite{KGPLC}. Thus, a
contextual HV model that gives the QM predictions must obey a
knowledge \emph{imbalance} principle:

\begin{quote}
  If one has maximal knowledge, then for every system, at every time,
  the amount of knowledge one possesses about the ontic state of the
  system at that time \emph{cannot exceed} the amount of knowledge one
  lacks.
\end{quote}

It is the author's firm belief that quantum contextuality is the main
cause of the difficulty of understanding QM, and that it plays an
important role in Quantum Information Theory. This note (and
\cite{KGPLC}) provides a link between information theoretical concepts
on the one side and quantum contextuality and the Kochen-Specker
theorem on the other.  There is a definite need to explore the
connection further.

\begin{theacknowledgments}
  The author would like to thank Adan Cabello, Matthias Kleinmann,
  Otfried G\"uhne, Jose Portillo, and Chris Fuchs for a number of
  interesting discussions.
\end{theacknowledgments}

\bibliographystyle{aipproc}   
\bibliography{Vaxjo2011}

\begin{thebibliography}{40}
\expandafter\ifx\csname natexlab\endcsname\relax\def\natexlab#1{#1}\fi
\providecommand{\enquote}[1]{``#1''}
\expandafter\ifx\csname url\endcsname\relax
  \def\url#1{\texttt{#1}}\fi
\expandafter\ifx\csname urlprefix\endcsname\relax\def\urlprefix{URL }\fi
\providecommand{\eprint}[2][]{\url{#2}}

\bibitem[Einstein et~al.(1935)]{Einstein1935777}
A.~Einstein, B.~Podolsky, and N.~Rosen, \emph{Physical Review} \textbf{47},
  777--780 (1935).

\bibitem[von Neumann(1931)]{VonNeumann31}
J.~von Neumann, \emph{Ann. of Math.} \textbf{32}, 191 (1931).

\bibitem[Bohr(1935)]{Bohr1935696}
N.~Bohr, \emph{Physical Review} \textbf{48}, 696--702 (1935).

\bibitem[Werner and Wolf(2001)]{WW01}
R.~F. Werner, and M.~M. Wolf, \emph{Quantum Information and Computation}
  \textbf{1}, 1--25 (2001).

\bibitem[Bohm and Hiley(1993)]{BH93}
D.~Bohm, and B.~J. Hiley, \emph{The Undivided Universe}, Routledge, London,
  1993.

\bibitem[Holland(1993)]{Holland93}
P.~R. Holland, \emph{The Quantum Theory of Motion}, Cambridge University Press,
  Cambridge, UK, 1993.

\bibitem[Bell(1964)]{Bell64}
J.~S. Bell, \emph{Physics (Long Island City, NY)} \textbf{1}, 195 (1964).

\bibitem[Aspect et~al.(1982)]{ADR82}
A.~Aspect, J.~Dalibard, and G.~Roger, \emph{Phys. Rev. Lett.} \textbf{49}, 1804
  (1982).

\bibitem[Weihs et~al.(1998)]{WJSWZ98}
G.~Weihs, T.~Jennewein, C.~Simon, H.~Weinfurter, and A.~Zeilinger, \emph{Phys.
  Rev. Lett.} \textbf{81}, 5039 (1998).

\bibitem[Rowe et~al.(2001)]{RKMSIMW01}
M.~A. Rowe, D.~Kielpinski, V.~Meyer, W.~M.~I. C.~A.~Sackett, C.~Monroe, and
  D.~J. Wineland, \emph{Nature (London)} \textbf{409}, 791 (2001).

\bibitem[Specker(1960)]{Specker60}
E.~Specker, \emph{Dialectica} \textbf{14}, 239 (1960), english version in
  \emph{The Logico-Algebraic Approach to Quantum Mechanics. Volume I:
  Historical Evolution}, edited by C. A. Hooker (Reidel, Dordrecht, Holland,
  1975), p.~135.

\bibitem[Bell(1966)]{Bell66}
J.~S. Bell, \emph{Rev. Mod. Phys.} \textbf{38}, 447 (1966).

\bibitem[Kochen and Specker(1967)]{KS67}
S.~Kochen, and E.~P. Specker, \emph{J. Math. Mech.} \textbf{17}, 59 (1967).

\bibitem[Roy and Singh(1993)]{RS93}
S.~M. Roy, and V.~Singh, \emph{Phys. Rev. A} \textbf{48}, 3379 (1993).

\bibitem[Cabello and Garc\'{\i}a-Alcaine(1998)]{CG98}
A.~Cabello, and G.~Garc\'{\i}a-Alcaine, \emph{Phys. Rev. Lett.} \textbf{80},
  1797 (1998).

\bibitem[Simon et~al.(2000)]{SZWZ00}
C.~Simon, M.~\.Zukowski, H.~Weinfurter, and A.~Zeilinger,
  \emph{Phys.~Rev.~Lett.} \textbf{85}, 1783 (2000).

\bibitem[Simon et~al.(2001)]{SBZ01}
C.~Simon, \v{C}. Brukner, and A.~Zeilinger, \emph{Phys.~Rev.~Lett.}
  \textbf{86}, 4427 (2001).

\bibitem[Larsson(2002)]{Larsson02}
J.-{\AA}. Larsson, \emph{Europhys. Lett.} \textbf{58}, 799 (2002).

\bibitem[Cabello et~al.(2008)]{CFRH08}
A.~Cabello, S.~Filipp, H.~Rauch, and Y.~Hasegawa, \emph{Phys. Rev. Lett.}
  \textbf{100}, 130404 (2008).

\bibitem[Klyachko et~al.(2008)]{KCBS08}
A.~A. Klyachko, M.~A. Can, S.~Binicio\u{g}lu, and A.~S. Shumovsky, \emph{Phys.
  Rev. Lett.} \textbf{101}, 020403 (2008).

\bibitem[Cabello(2008)]{Cabello08}
A.~Cabello, \emph{Phys. Rev. Lett.} \textbf{101}, 210401 (2008).

\bibitem[Badzi{\c a}g et~al.(2009)]{BBCP09}
P.~Badzi{\c a}g, I.~Bengtsson, A.~Cabello, and I.~Pitowsky, \emph{Phys. Rev.
  Lett.} \textbf{103}, 050401 (2009).

\bibitem[Meyer(1999)]{Meyer99}
D.~A. Meyer, \emph{Phys. Rev. Lett.} \textbf{83}, 3751 (1999).

\bibitem[Kent(1999)]{Kent99}
A.~Kent, \emph{Phys. Rev. Lett.} \textbf{83}, 3755 (1999).

\bibitem[Mermin(1999)]{Mermin99}
N.~D. Mermin, A {K}ochen-{S}pecker theorem for imprecisely specified
  measurements (1999), quant-ph/9912081.

\bibitem[Clifton and Kent(2000)]{CK00}
R.~Clifton, and A.~Kent, \emph{Proc. R. Soc. London, Ser. A} \textbf{456}, 2101
  (2000).

\bibitem[Havlicek et~al.(2001)]{HKSS01}
H.~Havlicek, G.~Krenn, J.~Summhammer, and K.~Svozil, \emph{J. Phys.~A}
  \textbf{34}, 3071 (2001).

\bibitem[Appleby(2002)]{Appleby02}
D.~M. Appleby, \emph{Phys. Rev.~A} \textbf{65}, 022105 (2002).

\bibitem[Cabello(2002)]{Cabello02}
A.~Cabello, \emph{Phys. Rev.~A} \textbf{65}, 052101 (2002).

\bibitem[Breuer(2002{\natexlab{a}})]{Breuer02a}
T.~Breuer, \enquote{A {K}ochen-{S}pecker theorem for unsharp spin 1
  observables,} in \emph{Non-locality and Modality}, edited by T.~Placek, and
  J.~Butterfield, Kluwer Academic, Dordrecht, Holland, 2002{\natexlab{a}}, p.
  195.

\bibitem[Breuer(2002{\natexlab{b}})]{Breuer02b}
T.~Breuer, \emph{Phys. Rev. Lett.} \textbf{88}, 240402 (2002{\natexlab{b}}).

\bibitem[Barrett and Kent(2004)]{BK04}
J.~Barrett, and A.~Kent, \emph{Stud. Hist. Phil. Sci. Part B: Stud. Hist.
  Philos. Mod. Phys.} \textbf{35}, 151 (2004).

\bibitem[La~Cour(2009)]{LaCour09a}
B.~R. La~Cour, \emph{Phys. Rev. A} \textbf{79}, 012102 (2009).

\bibitem[Cabello and Larsson(2010)]{CL10}
A.~Cabello, and J.-{\AA}. Larsson, \emph{Physics Letters A} \textbf{375}, 99
  (2010).

\bibitem[Spekkens(2007)]{Spekkens07}
R.~Spekkens, \emph{Phys. Rev. A} \textbf{75}, 032110 (2007).

\bibitem[Mermin(1990)]{Mermin90}
N.~D. Mermin, \emph{Phys. Rev. Lett.} \textbf{65}, 3373 (1990).

\bibitem[Peres(1990)]{Peres90}
A.~Peres, \emph{Phys. Lett. A} \textbf{151}, 107--108 (1990).

\bibitem[Larsson et~al.(2011)]{Larsson2011401}
J.-Å. Larsson, M.~Kleinmann, O.~Gühne, and A.~Cabello, \enquote{Violating
  noncontextual realism through sequential measurements,} in \emph{Advances In
  Quantum Theory: Proceedings Of The International Conference On Advances In
  Quantum Theory}, 2011, vol. 1327 of \emph{AIP Conference Proceedings}, pp.
  401--409.

\bibitem[Kleinmann et~al.(2011)]{KGPLC}
M.~Kleinmann, O.~G\"uhne, J.~R. Portillo, J.-{\AA}. Larsson, and A.~Cabello,
  \emph{New Journal of Physics} \textbf{13}, 113011 (2011).

\bibitem[Cabello(2010)]{Cabello10}
A.~Cabello, \emph{Phys. Rev. A} \textbf{82}, 032110 (2010).

\end{thebibliography}

\end{document}